\documentclass[prl,nofootinbib,superscriptaddress,twocolumn]{revtex4-1}
\usepackage[a4paper,left=1.5cm,right=1.5cm,top=3cm,bottom=3cm]{geometry}

\usepackage{verbatim}
\usepackage{color}
\usepackage{amsfonts,amssymb,mathrsfs,amsmath}
\usepackage{framed}
\usepackage{mdframed}
\usepackage{latexsym}
\usepackage{graphicx}
\usepackage{datetime}
\newdateformat{mydate}{\THEDAY{ }\monthname[\THEMONTH]{ }\THEYEAR}

\allowdisplaybreaks

\usepackage{tikz}
\usepackage{color}
\usepackage{framed}
\usepackage{hyperref}
\hypersetup{colorlinks, citecolor=bluscuro, linkcolor=black, urlcolor=bluscuro}
\definecolor{rossos}{cmyk}{0,1,1,0.55}
\definecolor{bluscuro}{rgb}{0.15, 0.2, .85}
\definecolor{bluchiaro}{cmyk}{1,.3,0.,0.1}

\newcommand{\be}{\begin{equation}}
\newcommand{\ee}{\end{equation}}
\renewcommand{\d}{{\rm d}}

\def\PBH{\text{\tiny PBH}}
\newcommand{\llp}{\left [}
\newcommand{\rrp}{\right ]}
\newcommand{\lp}{\left (}
\newcommand{\rp}{\right )}

\newcommand{\BH}{\textrm{\tiny BH}}
\newcommand{\s}{\textrm{\tiny S}}

\def\lsim{\mathrel{\rlap{\lower4pt\hbox{\hskip0.5pt$\sim$}}
    \raise1pt\hbox{$<$}}}         
\def\gsim{\mathrel{\rlap{\lower4pt\hbox{\hskip0.5pt$\sim$}}
    \raise1pt\hbox{$>$}}}         

\newcommand{\arXiv}[2]{\href{http://arxiv.org/pdf/#1}{{\tt [#2/#1]}}}
\newcommand{\arXivold}[1]{\href{http://arxiv.org/pdf/#1}{{\tt [#1]}}}

\begin{document}

\title{Standard Model Baryon Number Violation 
Seeded by Black Holes}

\author{V. De Luca}
\email{Valerio.DeLuca@unige.ch}
\address{D\'epartement de Physique Th\'eorique and Centre for Astroparticle Physics (CAP), Universit\'e de Gen\`eve, 24 quai E. Ansermet, CH-1211 Geneva, Switzerland}

\author{G. Franciolini}
\email{Gabriele.Franciolini@unige.ch}
\address{D\'epartement de Physique Th\'eorique and Centre for Astroparticle Physics (CAP), Universit\'e de Gen\`eve, 24 quai E. Ansermet, CH-1211 Geneva, Switzerland}

\author{A. Kehagias}
\email{Kehagias@central.ntua.gr}
\address{Physics Division, National Technical University of Athens, Zografou, Athens, 15780, Greece}

\author{A.~Riotto}
\email{Antonio.Riotto@unige.ch}
\address{D\'epartement de Physique Th\'eorique and Centre for Astroparticle Physics (CAP), Universit\'e de Gen\`eve, 24 quai E. Ansermet, CH-1211 Geneva, Switzerland}

\date{\today}

\begin{abstract}
\noindent
We show that black holes  with a Schwarzschild radius of the order of the electroweak scale may act as  seeds for the  baryon number violation within the Standard model via
 sphaleron transitions. The corresponding rate is    faster than the one in the pure vacuum and baryon number violation around black holes can take place during the evolution of the universe  after the electroweak phase transition.  We show however that this does not pose any threat for a  pre-existing baryon asymmetry in the universe.

\end{abstract}

\maketitle

\paragraph{Introduction.}
\noindent
It is well-known that, within the Standard Model (SM) of electroweak interactions,  the  baryon  ($B)$ and the lepton ($L$) symmetries are accidental and it is not possible to violate their corresponding charges at any order of perturbation theory.  Nevertheless,  non-perturbative effects  may give rise to processes which violate the baryon and the lepton  numbers. Indeed,  the presence of the non-abelian  group $SU(2)_L$ within the SM gauge group implies that the ground state is the sum of  an  infinite number of vacua which are classically degenerate and have different baryon (and lepton)  numbers. Static configurations, called sphalerons \cite{km},   corresponding to unstable solutions of the equations of motion and  to saddle points of the energy functional,  interpolate between two nearby vacua.  

The probability of  baryon number violation to occur  in the vacuum  through sphaleron transitions is  exponentially suppressed \cite{thooft}
\be
\Gamma_B\sim e^{-4\pi/\alpha_{\textrm{\tiny W} }}\sim e^{-150},
\ee
where $\alpha_{\textrm{\tiny W} }=g_2^2/4\pi$ is the $SU(2)_L$ gauge coupling constant.  
Such an exponential factor is  interpreted as the probability of making a transition from one classical vacuum to the closest one by quantum tunneling,  going through a
barrier of  energy $E_{\textrm{\tiny sph}}\sim 10$ TeV thanks to the formation of a sphaleron.
In more extreme situations,  like the primordial Universe, baryon and lepton number violation processes may be however  faster through classical transitions induced by the
 high-temperature environment and   play a significant role in the generation of the baryon asymmetry \cite{baureview}.

There are also arguments suggesting that all global symmetries, including the baryon one,  are violated when including  gravity \cite{arg}. In particular,  no-hair theorems tell us that global charges are swallowed by Black Holes (BHs). Indeed, quanta with global charge may scatter with a BH, leaving behind a BH with a slightly larger mass,  but  indeterminate global charge as dictated by the no-hair theorem. At the level of effective field theory,  one can imagine to integrate out  virtual BH  states  of mass $M_{\textrm{\tiny BH}}$ arising from quantum gravity,  leading to higher-dimensional  baryon number  violating operators suppressed by powers of $M_{\textrm{\tiny BH}}$, where $M_{\textrm{\tiny BH}}$  might be as small as the Planck mass $M_{\textrm{\tiny Pl}}$. 

What about baryon number violation induced by sphaleron transitions in the presence of BHs? In general, tunneling processes  may  be catalysed by the presence of impurities. A BH is a  gravitational impurity and indeed it has been shown that  BHs can trigger electroweak SM vacuum instability in their vicinity, both at zero temperature \cite{i1,i2, Gregory:2018bdt, Gregory:2020hia} and in the early universe \cite{Tetradis:2016vqb, Gorbunov:2017fhq, i3,Mukaida:2017bgd, Kohri:2017ybt, i4},  and baryon number violations through interactions with skyrmions~\cite{Luckock:1986tr, Moss:2000hf}.

Since we are dealing with SM sphaleron configurations,  a simple  estimate tells us that the typical Schwarzschild radius of the BH able 
 to alter the rate of the baryon number violation is
 \be
r_{\textrm{\tiny S}}=2GM_{\textrm{\tiny BH} }\sim \frac{1}{M_{\textrm{\tiny W} }}\sim \frac{1}{g_2 v},
\ee
where $G=1/M_{\textrm{\tiny Pl}}^2$ and $v=246$ GeV is the Vacuum Expectation Value (VEV) of the Higgs field. This leads to BH masses in the ballpark of 
\be
 \label{estimate}
M_{\textrm{\tiny BH} }= {\cal O}(1) \cdot  10^{-22} M_\odot\sim 10^{17} M_{\textrm{\tiny Pl} },
\ee
i.e. to BHs which evaporate with a typical lifetime of ${\cal O}(1)$  yr and which might have been present during the evolution of the Universe. 

We are going to show that  baryon number violation through sphaleron transitions in the presence of  such BHs can be  faster than in the pure vacuum and we will offer as well some considerations about what may happen  should these tiny BHs be present during the evolution of the universe.

\vskip 0.5cm
\noindent
\paragraph{Baryon  number violation seeded by  BHs.}
To  study the  influence of BHs on the sphaleron transitions we start from the action of the Higgs doublet field $\phi$ along with a $SU(2)_L$  gauge field $W^a_\mu $ (including the abelian hypercharge group $U(1)_Y$ does not change our results) in a curved spacetime 
\begin{eqnarray}
\label{action}
S&=& \int_{\mathcal{M}} \d^4 x \sqrt{-g}  \left[  \frac{\mathcal{R}}{16 \pi G}  - g^{\mu \nu} (D_\mu \phi)^\dagger  D_\nu \phi - V(\phi) \right. \nonumber\\
&-& \left.\frac{1}{4} g^{\mu \rho} g^{\nu \sigma} F^a_{\rho \sigma} F^a_{\mu \nu}\right]+ \frac{1}{8\pi G} \int_{\partial \mathcal{M}}\mathcal{K} \sqrt{\gamma} \d^3 y,
\end{eqnarray}
where $V(\phi)$ 
is the Higgs potential and we have added   the Gibbons-Hawking-York boundary term as we deal with  a spacetime manifold $\mathcal{M}$ with a BH horizon. The spacetime geometry around the BH can be taken static and spherically symmetric, such that its metric takes a Schwarzschild-like form
\begin{eqnarray}
\d s^2 &=& -e^{2 \delta (r)} A (r) \d t^2 + A^{-1} (r) \d r^2 + r^2  \d \Omega^2,\nonumber\\
A(r) &=& 1 - \frac{2 G M (r)}{r},
\end{eqnarray}
where   $A (r)$ vanishes at the horizon 
\be
r_{\textrm{\tiny S}} \equiv 2 G M (r_{\textrm{\tiny S}})=2 G M_{\textrm{\tiny BH}}.
\ee
 A suitable ansatz for the gauge and Higgs field is \cite{km} 
\begin{eqnarray}
W_i^a\sigma^a\d x^i&=&-\frac{2i}{g_2}f(g_2 v r)\d U^\infty(U^\infty)^{-1},\nonumber\\
\phi&=&\frac{v}{\sqrt{2}} h(g_2 v r) U^\infty\left(\begin{array}{c}
0\\
1\end{array}\right),
\end{eqnarray}
where
\be
U^\infty=\frac{1}{r}\left(
\begin{array}{cc}
z& x+ iy\\
-x+i y & z\end{array}\right).
\ee
Since we are ultimately interested in the energy functional, we perform an analytical continuation of the action to the Euclidean metric with $t = i \tau$, taking $\tau$ to be periodic with period $1/T$ (to be identified with the relevant temperature of the system). 

By setting $\xi=g_2 v r$ and  expanding the mass with respect to its value at the horizon  
\be
M (\xi) = M_\BH + \delta M (\xi),
\ee
we can write the equations of motion
\begin{eqnarray}
\label{EOM}
\delta M' &=& \frac{4 \pi v}{g_2} \left[ \frac{\xi^2}{2} A (h')^2 + \epsilon \frac{\xi^2}{4} (h^2 -1)^2  + 4 A (f')^2\right.\nonumber\\
&+&\left. \frac{8}{\xi^2} f^2 (1-f)^2 + \xi^2 \mathcal{F}(\xi) \tilde \epsilon h^2 + h^2(1-f)^2 \right],  \nonumber \\
\delta' &=& 2 \alpha^2 \llp \frac{8}{\xi} (f')^2 +  \xi (h')^2 \rrp , \nonumber \\
(A e^\delta f')' & = &\frac{2}{\xi^2} e^\delta f (1-f) (1-2 f) -  \frac{1}{4}e^\delta h^2 (1- f),  \nonumber \\
(\xi^2 A e^\delta h')' & =& \epsilon \xi^2 e^\delta h (h^2 - 1) + 2 e^\delta h (1-f)^2\nonumber\\
&+&2  e^\delta \xi^2   \mathcal{F}(\xi)  \tilde \epsilon  h, 
\end{eqnarray}
where $\alpha = \sqrt{4 \pi G}(v/\sqrt{2})$, $\epsilon = \lambda/g_2^2$ and  $\tilde \epsilon = \tilde \lambda/768 \pi^2$. Here $\lambda$ is the quartic coupling of the Higgs and we have 
written the Higgs potential as 
\be
V(h,\xi) \simeq \frac{\lambda}{4} v^4 (h^2-1)^2 + \frac{\tilde \lambda g_2^2}{768 \pi^2} \mathcal{F}(\xi) v^4 h^2.
\ee
The second term is due to the  vacuum polarization effect of the Hawking radiation  originating at one-loop from the interactions of the Higgs with the other SM particles in the vicinity of the horizon of the BH~\cite{Hayashi:2020ocn}. This term is  very similar to the
finite temperature correction to the mass squared of the Higgs $\sim T^2h^2$ in a plasma at finite temperature $T$. The key difference is that the effective temperature depends on the distance from the horizon  \cite{candelas, Moss:1984zf} (being $\xi_\s \equiv g_2 v r_\s$  the dimensionless BH horizon)
\be
\mathcal{F}(\xi) =   
\begin{cases}
\frac{3}{4\xi_\s^2} & \xi \simeq \xi_\s, \\
     \frac{1}{\xi^2} & \xi \gg  \xi_\s,
    \end{cases}    
\ee
so that, close to the horizon, the correction to the potential acquires the familiar  form $T_{\textrm{\tiny H}}^2 h^2$, where \cite{Hawking:1974sw}
\begin{equation}
\label{Hawking}
T_{\textrm{\tiny H}}= \frac{1}{8 \pi G M_\BH} \simeq 10  \lp \frac{M_\BH}{5\cdot 10^{-22}\,M_\odot}\rp^{-1} \text{GeV}
\end{equation}
is the Hawking temperature. We adopt here the Unruh vacuum \cite{Unruh:1976db} as the most appropriate vacuum for our physical situation. Indeed, in the following we will consider the case in which  the temperature of the universe is different from the Hawking temperature. As such,  the Hartle-Hawking  vacuum \cite{Hartle:1976tp} is not the proper one as it assumes full and static thermal equilibrium with the surrounding plasma. 

The  effective coupling $\tilde \lambda$  is given by 
\be
\tilde \lambda = 24 \lp \frac{3}{16}g_2^2 + \frac{1}{16}g_1^2 + \frac{1}{4}y_t^2 + \frac{\lambda}{2}\rp\sim 9.56,
\ee
computed in terms of the $g_2$, $g_1$ (the gauge coupling of the $U(1)_Y$ group), and top Yukawa coupling $y_t$, all evaluated at the electroweak scale~\cite{Buttazzo:2013uya}. 

 Since $\alpha\ll 1$, one can approximate $\delta' \simeq 0$ and, given that the metric has to approach the Minkowski spacetime at infinity, the leading order solution of Eq.~\eqref{EOM} gives  $\delta \simeq 0$. 
The equations for the  gauge and Higgs fields then simplify to
\begin{align}
	\label{EOM2}
		f'' + \frac{\xi_\s}{\xi (\xi-\xi_\s)} f'& = \frac{2}{\xi(\xi-\xi_\s)} f (1-f) (1-2 f) \nonumber\\
	&-  \frac{1}{4} \frac{\xi}{\xi-\xi_\s} h^2 (1- f) ,
	\nonumber \\
	h'' \!+\!\lp \frac{2}{\xi} \!+\! \frac{\xi_\s}{\xi(\xi\!-\!\xi_\s)} \rp h'  &= \frac{\xi}{\xi-\xi_\s}\epsilon h (h^2\!-\!1)+\frac{2 \xi \mathcal{F}(\xi)}{(\xi \!-\! \xi_h)} \tilde \epsilon h \nonumber\\
	& + \frac{2}{\xi(\xi\!-\!\xi_\s)} h (1\!-\!f)^2.
\end{align}
\begin{figure*}[t!]
	\centering
	\includegraphics[width=.49 \linewidth]{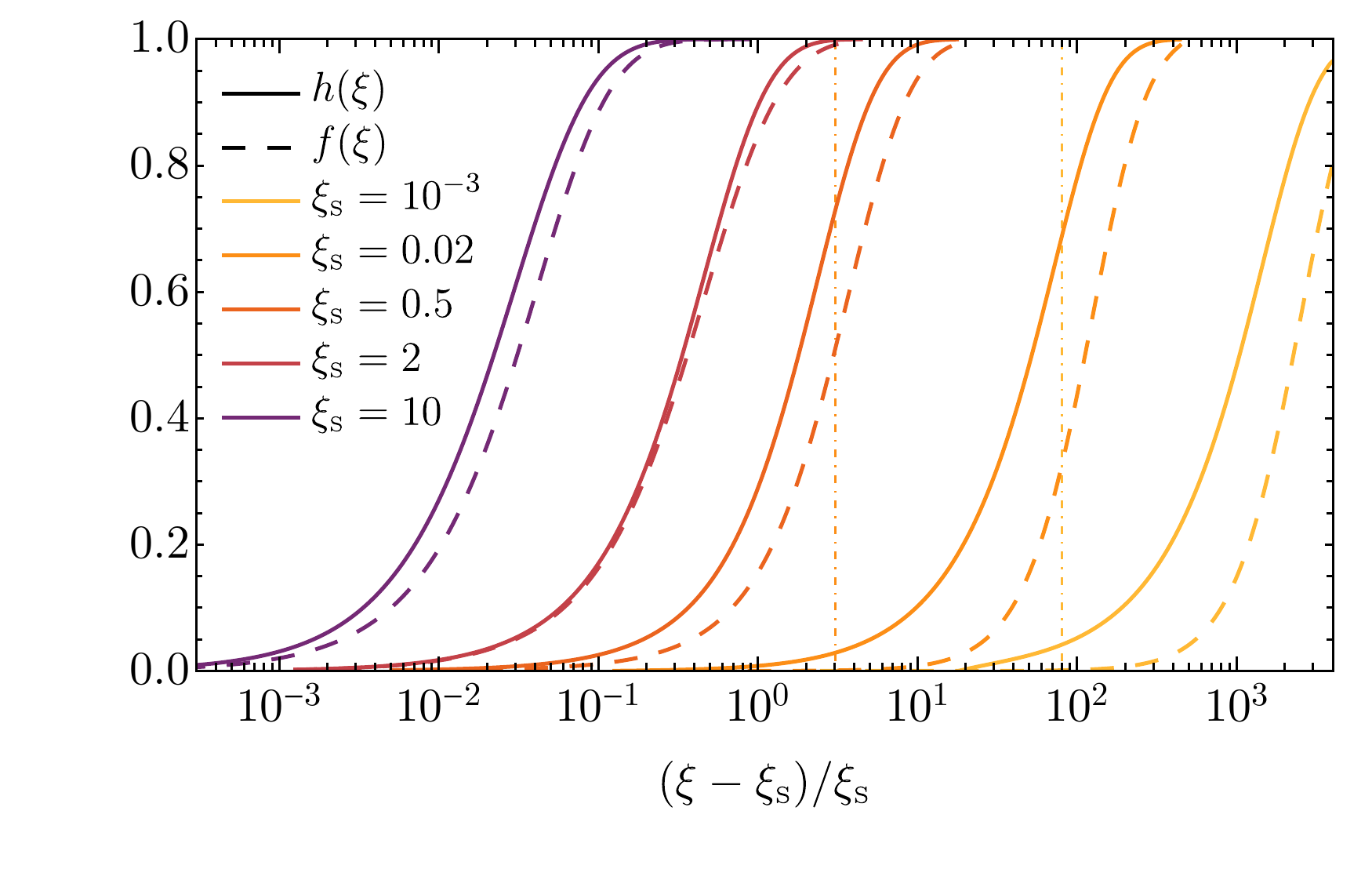}	
	\includegraphics[width=.49 \linewidth]{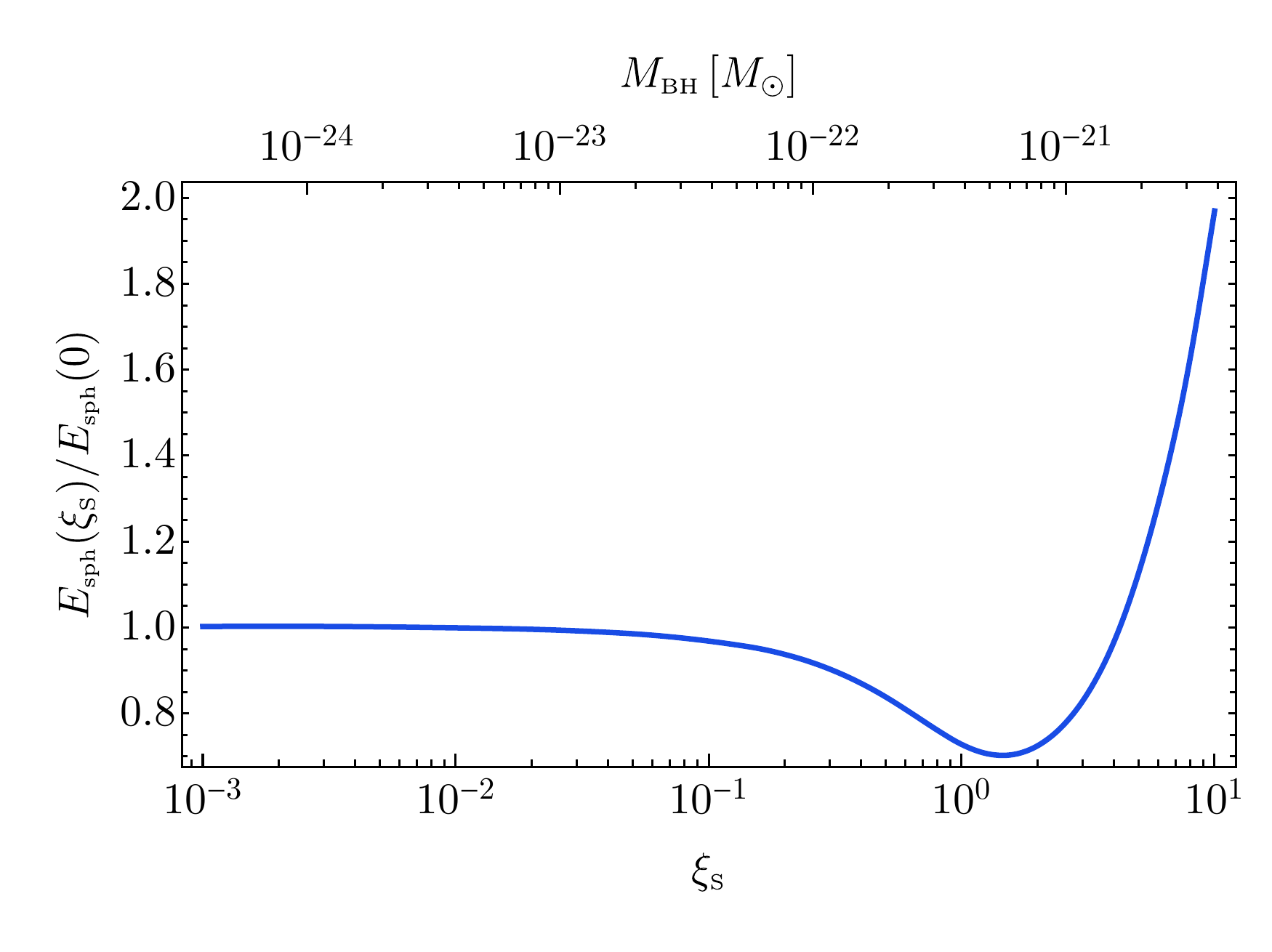}
	\caption{\small  {\it Left}:  Behaviour of the Higgs and $SU(2)_L$ fields in terms of the radial coordinate for different values of the rescaled horizon $\xi_\s$. The vertical lines indicate the radial position where the Hawking temperature is able to restore the symmetry.
	 {\it Right}: 
	Values of the rescaled sphaleron energy for different BH masses.
	}
	\label{figKli1}
\end{figure*}
In order to solve the equations of motion,  we have to impose proper boundary conditions.
At infinity the metric has to approach the Minkowski spacetime and the fields have to be in their true vacuum, 
\be
  f(\xi) \to 1, ~~ h(\xi) \to 1, ~~ {\rm when}~~\xi \to \infty.
\ee
At the BH horizon $\xi \to \xi_\s$, one can impose the boundary conditions setting the fields in the false vacuum
\be
 f(\xi) \to 0, ~~ h(\xi) \to 0, ~~  {\rm when}~~ \xi \to \xi_\s. 
\ee
The numerical solutions of the equations of motion  can be found in the left panel of Fig.~\ref{figKli1} for different rescaled BH horizons. For small enough BHs, there exists a critical radius below which the vacuum polarization effect induced by the Hawking radiation leads to the restoration of the symmetry close to the horizon, nevertheless allowing for a sphaleron solution interpolating between the unbroken and broken phase.

The  characteristic mass contribution at infinity is 
\begin{align}
\label{energy}
\delta M_\infty &=   \frac{4 \pi v}{g_2} \int_{\xi_\s}^\infty \d \xi \left[ \frac{1}{2} \xi (\xi-\xi_\s)  (h')^2 + \epsilon \frac{\xi^2}{4} (h^2 -1)^2\right. \nonumber\\
	  &\hspace{1.5cm}\left. +\frac{4}{\xi}(\xi-\xi_\s) (f')^2 + \frac{8}{\xi^2} f^2 (1-f)^2  
	 \right. \nonumber \\	  
	  &\hspace{1.5cm} \left.+\xi^2 \mathcal{F}(\xi) \tilde \epsilon h^2+ h^2(1-f)^2 \right.\Big], 
\end{align}
which  has to be thought as the sphaleron energy in the presence of a BH, 
\be
\delta M_\infty=E_\text{\tiny sph}(M_\BH).
\ee
 Indeed,  in the limit of flat spacetime with no BH $(r_\s\ll 1/g_2 v$),   we get 
 \be
 E_\text{\tiny sph}(0)\simeq 1.92 \frac{4\pi v}{g_2}, 
 \ee 
 which reproduces the standard result for the current physical mass of the Higgs (i.e. for $\epsilon\simeq 0.3$), see right panel of Fig.~\ref{figKli1}. Notice that the effect of the vacuum polarization in the Higgs potential, in the limit of tiny BH masses, is minor because the radius of the sphaleron configuration is located away from the Schwarzschild radius.

For small  BH masses  the sphaleron radius is large compared to the  Schwarzschild radius and its energy is only slightly perturbed compared to the vacuum solution. As the seed BH masses increase, the sphaleron radius approaches the horizon and the BH helps catalysing the sphaleron transitions. For larger BH masses, it is energetically more costly to  generate the sphaleron solution as its  characteristic size is required to be larger than the BH horizon and therefore  larger  than $\sim 1/g_2 v$.
Notice also that the minimum BH mass is consistent
with the estimate (\ref{estimate}).

\begin{figure*}[t!]
	\centering
	\includegraphics[width=.49 \linewidth]{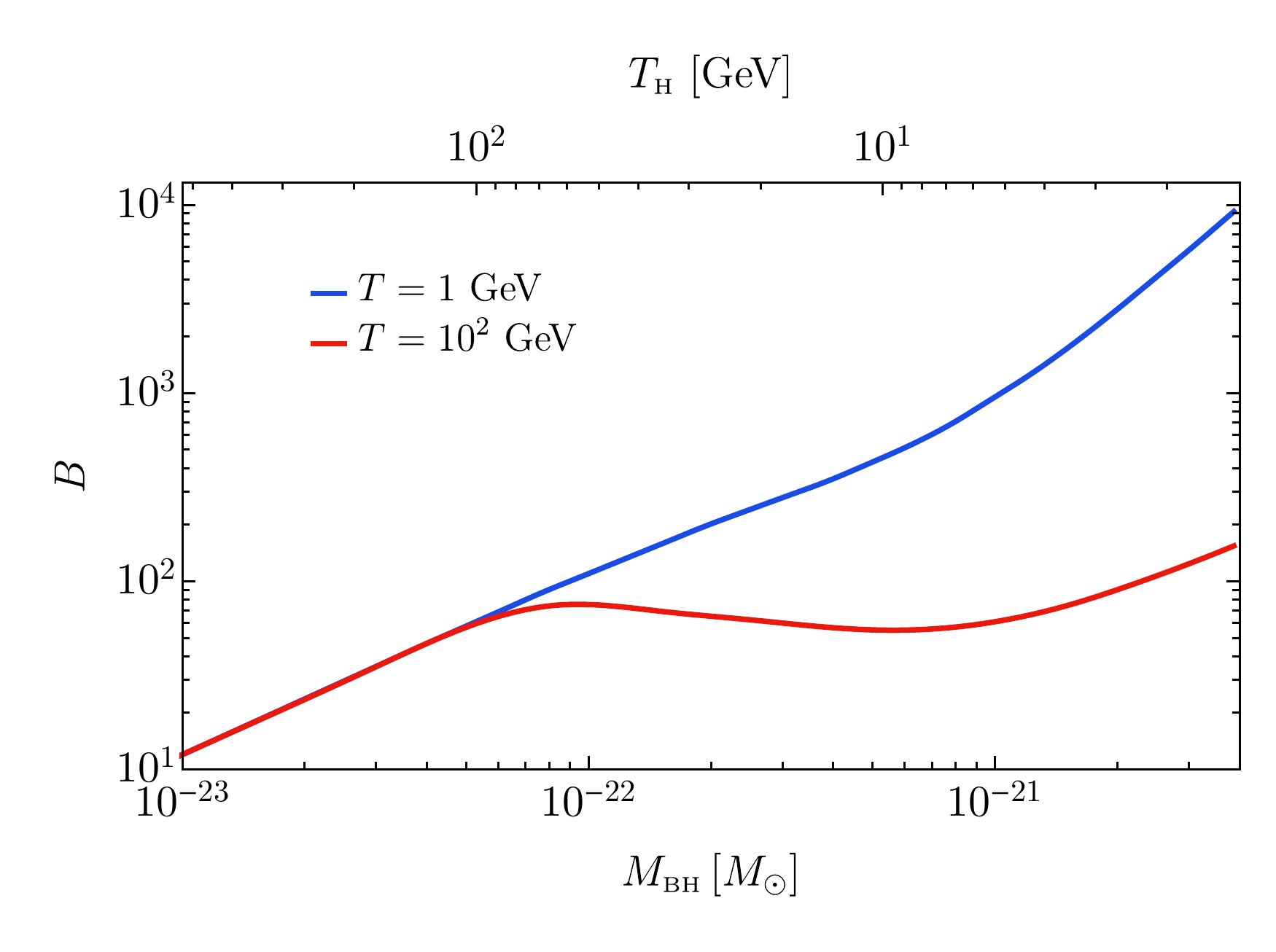}	
	\includegraphics[width=.49 \linewidth]{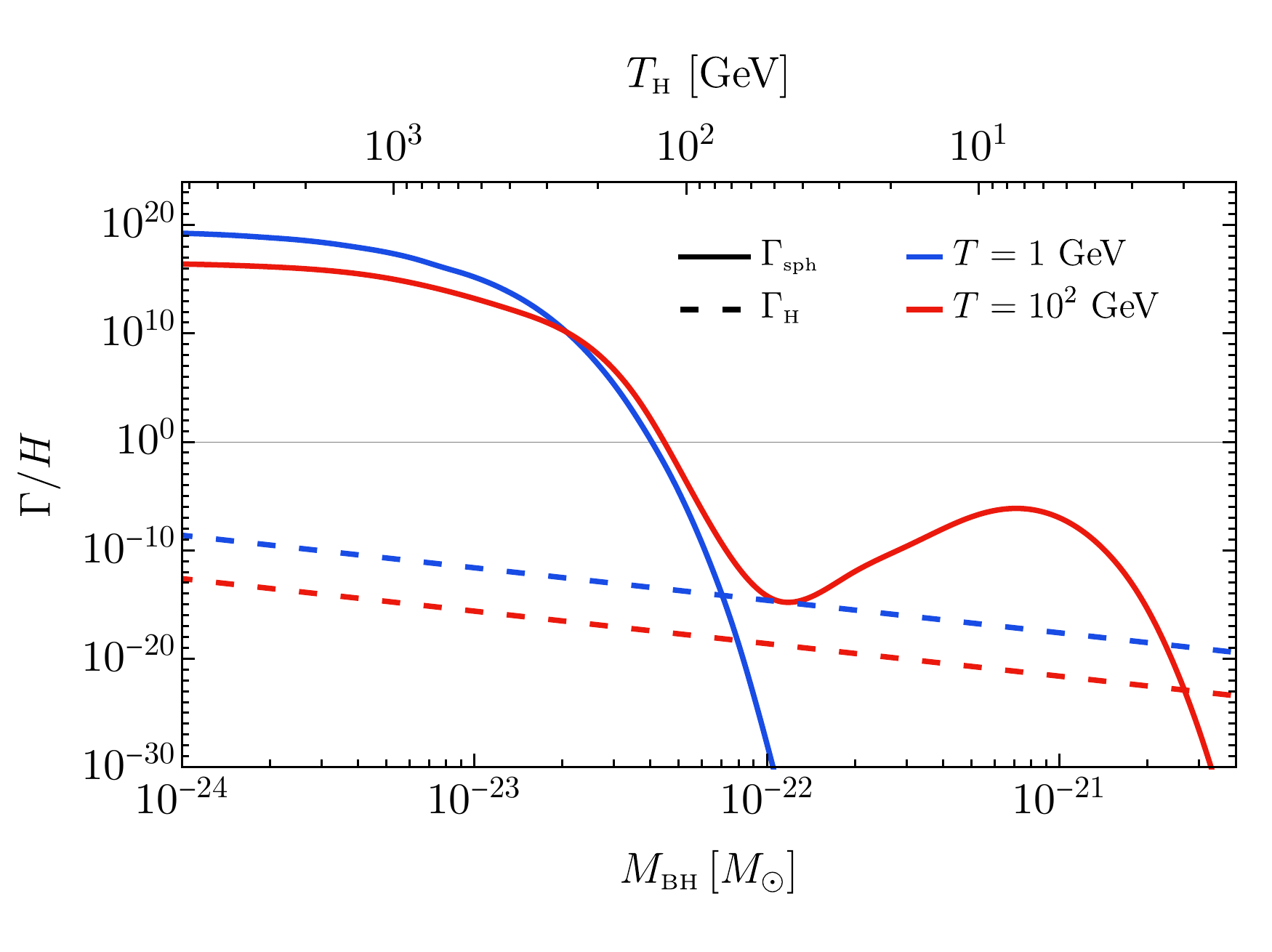}
	\caption{\small {\it Left}: Bounce exponent in terms of the BH mass (and corresponding Hawking temperature) for different temperatures of the plasma.  {\it Right}: Ratio between the sphaleron rate in the presence of a BH of mass $M_\text{\tiny BH}$ and  Hubble rate. For comparison, we also show the evaporation rate. Notice that for small masses the sphaleron rate decreases as $\sqrt{B}$, while for larger masses it has a bump due to the plasma temperature effect.}
	\label{figB}
\end{figure*}

\paragraph{Rate of baryon  number violation seeded by  BHs.}
How fast can the  baryon number violation  take place in the vicinity of a BH? The vacuum decay rate takes the form~\cite{Gregory:2013hja}
\be
\label{rate}
\Gamma_\text{\tiny sph}(M_\text{\tiny BH}) \sim \sqrt{\frac{B}{2\pi}} \frac{1}{\ell_{\textrm{\tiny sph}}} e^{-B},
\ee
where $\ell_\text{\tiny sph}$ is the size  of the sphaleron configuration. For large BH masses, it turns out to be comparable to the   Schwarzschild radius $r_\s$, while for
small BH masses it is of the order of $1/g_2 v$. 
The dimensionless term $\sqrt{B/2\pi}$ comes from the normalization of the zero mode associated with time translation symmetry, in terms of 
the exponent $B$  given by the difference between the Euclidean action of the bounce solution and the one before the transition. For a static solution this coincides with the difference between the BH areas at the horizon and at infinity.  As one can easily check, 
for the static  solution the bulk part of the energy functional vanishes due to the Hamiltonian constraint, while the boundary terms give the BH Bekenstein-Hawking entropy at the BH horizon~\cite{Gregory:2013hja}
\be
B = - \frac{\mathcal{A}_\s}{4 G} +  \frac{\mathcal{A}_\infty}{4 G} = 4 \pi G \llp M^2_\infty - M^2_\text{\tiny BH} \rrp.
\ee
A nice and useful interpretation of this formula may be obtained by expanding at the leading order $M_\infty=M_\BH+\delta M_\infty$ (we have checked this approximation to be valid in the BH mass range we are concerned with). One obtains
\be
\label{interp}
B \simeq   8 \pi G M_\text{\tiny BH}  \delta M_\infty \simeq \frac{ E_\text{\tiny sph}(M_\BH)}{T_\text{\tiny H}},
\ee
where in the last passage we have recognised the BH temperature $T_\text{\tiny H} = 1/8 \pi G M_\text{\tiny BH}$. The expression (\ref{interp}) tells us that the exponential factor ${\rm exp}(-B)$  for the
sphaleron transition may be thought of as the standard Boltzmann suppression factor in a thermal environment where the temperature of the system is indeed the Hawking temperature. This interpretation allows as well to smoothly interpolate between the zero  and the finite temperature limits. In the case in which the BH is immersed into
a plasma at finite temperature  $T$, as in the case of the Primordial BHs (PBHs), the sphaleron baryon number violating rate  is expected to go as  $ {\rm exp}(-E_\text{\tiny sph}(M_\BH)/T)$ for $T\gsim T_\text{\tiny H}$, fitting  the exponential factor (\ref{rate}) at $T\sim T_\text{\tiny H}$. In such limit one must use the VEV  of the Higgs field at finite temperature $v(T)$, see Refs.~\cite{Quiros:1999jp,Riotto:1998bt}.

In Fig. \ref{figB} we have plotted, on the left panel, the bounce factor $B$ as a function of the BH mass and for two choices of the plasma temperature. One can see that, for enough small   BH masses,   the bounce   $B$ can be much smaller than the vacuum bounce $4\pi/\alpha_{\textrm{\tiny W}}\sim 150$, reaching values of order unity for masses $M_\BH \sim 10^{-24} M_\odot $, below which the validity of  the computation breaks down. This happens in the region where the expression (\ref{interp}) applies. Moreover, as the BH masses increases, the finite temperature of the thermal bath dominates over the Hawking temperature, leading to a suppression of the bounce exponent.

\vskip 0.5cm
\noindent
\paragraph{Some further considerations.}  PBHs with masses of the order of $10^{-22} M_\odot$ may have populated the early Universe, even if with an abundance, normalized to the dark matter one,  $f_\PBH = \Omega_\PBH/ \Omega_\text{\tiny DM}\lsim 10^{-4}$ to avoid bounds from Big Bang nucleosynthesis \cite{Josan:2009qn, bound}.

 If they are formed by the collapse of large overdensities within a horizon volume, their formation temperature is~\cite{sasaki}
\be
T_\text{\tiny f}\simeq 10^{10}    \left( \frac{M_\BH}{5\cdot 10^{-22}\,M_\odot}\right)^{-1/2}\, {\rm GeV},
\ee
while their Hawking temperature is in Eq.~\eqref{Hawking}.
The light PBHs we are concerned with are always born with a Hawking temperature which is smaller than the plasma temperature. The rate of evaporation for the masses under consideration is  quite small and given by \cite{page}
\be
\Gamma_{\textrm{\tiny H}}  \sim 4\cdot  10^{-33} \left(\frac{10^{-22} M_\odot}{M_\BH}\right)^{-3}\,{\rm GeV}.
\ee
In  first approximation, one may consider the PBH masses as constant in time for our considerations. 
A comparison between the baryon number violation rate and the evaporation rate, both in terms of the Hubble rate, can be found in the right panel of Fig. \ref{figB}. 
The evaporation rate becomes relevant only for BH masses smaller than $10^{-28} M_\odot$, for which evaporation is effective at temperatures around $100 \, {\rm GeV}$.

Now, at very high temperatures thermal fluctuations induce unsuppressed
baryon number violation through sphaleron transitions till the electroweak phase transition takes place \cite{baureview}. In the SM this happens at $T_{\textrm{\tiny EW}}\simeq 163$ GeV for the current mass of the Higgs. At smaller temperatures and away from the PBHs,  the sphalerons are inactive and baryon number violation is suppressed by the exponential ${\rm exp}(-E_\text{\tiny sph}(0)/T)$. However, even after the electroweak phase transition, 
   baryon number violation can take place at a rate faster than the rate of the expansion of the universe around  the PBHs, see Fig. \ref{figB} right panel,  where for each BH mass we have taken the maximum between the plasma temperature and the Hawking temperature to evaluate the suppression factor. 
   
   Does this represent a threat for the scenarios where the baryon asymmetry of the universe is generated before or at the electroweak phase transition? At the time of formation, the fraction of PBHs per horizon is given by \cite{sasaki}
\be
\beta(T_{\text{\tiny f}})\simeq  10^{-19}\left(\frac{M_\BH}{5\cdot 10^{-22} M_\odot}\right)^{1/2} f_\PBH.
\ee
 Big Bang nucleosynthesis bounds limit the PBH mass fraction at formation to be $\beta(T_{\text{\tiny f}}) \lesssim 10^{-23}$ for the range of masses of interest~\cite{Josan:2009qn}.
The number ${\cal N}$  of causally independent regions at a time during the radiation-dominated era with temperature $T$ and  currently within our horizon is given by
${\cal N}\sim 10^{34} (T/{\rm GeV})^3$. This means that the number density of PBHs at a given temperature $T$ normalized to the photon number density $n_\gamma$ is  approximately given by 
%
\be
\frac{n_\PBH}{n_\gamma} \frac{1}{\eta}\sim 10^{-34} \lp \frac{\eta}{10^{-9}} \rp^{-1}\left(\frac{M_\BH}{5\cdot 10^{-22} M_\odot}\right)^{-1} f_\PBH,
\ee
 where we have introduced the baryon asymmetry $\eta = n_\text{\tiny b}/n_\gamma $ normalised to the current constrained value~\cite{Riotto:1998bt}.
Luckily, the PBH  density is  too small to have any impact  on the pre-exisisting baryon asymmetry.
We believe that such conclusion would be hardly changed envisaging other scenarios of PBH formation in the early universe.

\vskip 0.5cm
\noindent
\paragraph{Conclusions.}  We have studied the violation of the baryon number within the SM induced by sphaleron transitions around a BH. Our findings indicate that the bounce for such transitions   may be much smaller than the one in the absence of BHs if their  Schwarzschild radius is of the order of the electroweak scale. Around PBHs the violation of the baryon number takes place at temperatures below the electroweak phase transition.  However our findings indicate that the baryon asymmetry of the universe is unlikely to be wiped out by the presence of  PBHs acting as seeds of the  sphaleron transitions.

\vskip 0.3cm
\noindent
\paragraph{Acknowledgments.}
\noindent
 We thank N. Tetradis for interesting discussions. V.DL., G.F. and 
A.R. are supported by the Swiss National Science Foundation 
(SNSF), project {\sl The Non-Gaussian Universe and Cosmological Symmetries}, project number: 200020-178787.

\bigskip



\begin{references}
\bibitem{km} R.F. Klinkhamer and N.S. Manton, Phys. Rev. {\bf D}30, 2212 (1984). 
\bibitem{thooft} G. 't Hooft, Phys. Rev. Lett. {\bf 37}, 37 (1976). 



\bibitem{baureview}
A.~Riotto and M.~Trodden,
Ann. Rev. Nucl. Part. Sci. \textbf{49}, 35-75 (1999)
\arXivold{hep-ph/9901362}.





\bibitem{arg} T.~Banks and N.~Seiberg,
Phys. Rev. D \textbf{83}, 084019 (2011)
\arXiv{1011.5120}{hep-th}.

\bibitem{i1} P.~Burda, R.~Gregory and I.~Moss,
Phys. Rev. Lett. \textbf{115}, 071303 (2015)
\arXiv{1501.04937}{hep-th}.

\bibitem{i2} P.~Burda, R.~Gregory and I.~Moss,
JHEP \textbf{08}, 114 (2015)
\arXiv{1503.07331}{hep-th}.

\bibitem{Gregory:2018bdt}
R.~Gregory, K.~M.~Marshall, F.~Michel and I.~G.~Moss,
Phys. Rev. D \textbf{98} (2018) no.8, 085017
\arXiv{1808.02305}{hep-th}.

\bibitem{Gregory:2020hia}
R.~Gregory, I.~G.~Moss, N.~Oshita and S.~Patrick,
JHEP \textbf{09} (2020), 135
\arXiv{2007.11428}{hep-th}.


\bibitem{Tetradis:2016vqb}
N.~Tetradis,
JCAP \textbf{09} (2016), 036
\arXiv{1606.04018}{hep-ph}.

\bibitem{Gorbunov:2017fhq}
D.~Gorbunov, D.~Levkov and A.~Panin,
JCAP \textbf{10} (2017), 016
\arXiv{1704.05399}{astro-ph.CO}.

\bibitem{i3} D.~Canko, I.~Gialamas, G.~Jelic-Cizmek, A.~Riotto and N.~Tetradis,
Eur. Phys. J. C \textbf{78}, no.4, 328 (2018)
\arXiv{1706.01364}{hep-th}.

\bibitem{Mukaida:2017bgd}
K.~Mukaida and M.~Yamada,
Phys. Rev. D \textbf{96} (2017) no.10, 103514
\arXiv{1706.04523}{hep-th}.

\bibitem{Kohri:2017ybt}
K.~Kohri and H.~Matsui,
Phys. Rev. D \textbf{98} (2018) no.12, 123509
\arXiv{1708.02138}{hep-ph}.

\bibitem{i4} D.~C.~Dai, R.~Gregory and D.~Stojkovic,
Phys. Rev. D \textbf{101}, no.12, 125012 (2020)
\arXiv{1909.00773}{hep-ph}.

\bibitem{Luckock:1986tr}
H.~Luckock and I.~Moss,
Phys. Lett. B \textbf{176} (1986), 341-345

\bibitem{Moss:2000hf}
I.~G.~Moss, N.~Shiiki and E.~Winstanley,
Class. Quant. Grav. \textbf{17} (2000), 4161-4174
\arXivold{gr-qc/0005007}.

\bibitem{Hayashi:2020ocn}
T.~Hayashi, K.~Kamada, N.~Oshita and J.~Yokoyama,
JHEP \textbf{08} (2020), 088
\arXiv{2005.12808}{hep-th}.

\bibitem{candelas}
P.~Candelas,
Phys. Rev. D \textbf{21},  2185 (1980).

\bibitem{Moss:1984zf}
I.~G.~Moss,
Phys. Rev. D \textbf{32} (1985), 1333.

\bibitem{Hawking:1974sw}
S.~W.~Hawking,
Commun. Math. Phys. \textbf{43} (1975), 199-220
[erratum: Commun. Math. Phys. \textbf{46} (1976), 206].

\bibitem{Unruh:1976db}
W.~G.~Unruh,
Phys. Rev. D \textbf{14} (1976), 870.

\bibitem{Hartle:1976tp}
J.~B.~Hartle and S.~W.~Hawking,
Phys. Rev. D \textbf{13} (1976), 2188-2203.

\bibitem{Buttazzo:2013uya}
D.~Buttazzo, G.~Degrassi, P.~P.~Giardino, G.~F.~Giudice, F.~Sala, A.~Salvio and A.~Strumia,
JHEP \textbf{12} (2013), 089
\arXiv{1307.3536}{hep-ph}.

\bibitem{Gregory:2013hja}
R.~Gregory, I.~G.~Moss and B.~Withers,
JHEP \textbf{03} (2014), 081
\arXiv{1401.0017}{hep-th}.

\bibitem{Riotto:1998bt}
A.~Riotto,
\arXivold{hep-ph/9807454}.

\bibitem{Quiros:1999jp}
M.~Quiros,
\arXivold{hep-ph/9901312}.

\bibitem{Josan:2009qn}
A.~S.~Josan, A.~M.~Green and K.~A.~Malik,
Phys. Rev. D \textbf{79} (2009), 103520
\arXiv{0903.3184}{astro-ph.CO}.


\bibitem{bound} 
B.~Carr, K.~Kohri, Y.~Sendouda and J.~Yokoyama,
\arXiv{2002.12778}{astro-ph.CO}.

\bibitem{sasaki} For a review, see 
M.~Sasaki, T.~Suyama, T.~Tanaka and S.~Yokoyama,
Class. Quant. Grav. \textbf{35}, no.6, 063001 (2018)
\arXiv{1801.05235}{astro-ph.CO}.

 \bibitem{page}
D.~N.~Page,
Phys. Rev. D \textbf{13} (1976), 198-206.


\end{references}
\end{document}